\documentclass[12pt]{iopart}

\usepackage{iopams}  

\usepackage{dsfont}
\usepackage{multirow}
\usepackage{makecell}
\usepackage{array}
\usepackage{graphicx}
\usepackage[colorlinks]{hyperref}
\usepackage{cite}
\usepackage[
singlelinecheck=false 
]{caption}

\newcommand{\angdots}[3][-20]{
  \rotatebox{#1}{\makebox[0pt]{\makebox[#2]{\xleaders\hbox{$\cdot$\hskip#3}\hfill\kern0pt}}}%
}

\newcommand\binom[2]{\left( #1 \atop  #2 \right)}
\newcommand\expval[1]{\langle #1 \rangle}

\newcommand{\p}{\mathrm{p}}
\newcommand{\bA}{{\bf A}}
\newcommand{\bF}{{\bf F}}
\newcommand{\bH}{{\bf H}}
\newcommand{\bK}{{\bf K}}
\newcommand{\bQ}{{\bf Q}}
\newcommand{\bS}{{\bf S}}
\newcommand{\bV}{{\bf V}}
\newcommand{\bW}{{\bf W}}
\newcommand\blue[1]{\textcolor{blue}{#1}}

\begin{document}

\title{Recursion for the smallest eigenvalue density of $\beta$-Wishart-Laguerre ensemble}

\author{Santosh Kumar}
\address{Shiv Nadar University, Gautam Buddha Nagar, Uttar Pradesh - 201314, India}
\ead{skumar.physics@gmail.com}

\begin{abstract}
The statistics of the smallest eigenvalue of Wishart-Laguerre ensemble is important from several perspectives. The smallest eigenvalue density is typically expressible in terms of determinants or Pfaffians. These results are of utmost significance in understanding the spectral behavior of Wishart-Laguerre ensembles and, among other things, unveil the underlying universality aspects in the asymptotic limits. However, obtaining exact and explicit expressions by expanding determinants or Pfaffians becomes impractical if large dimension matrices are involved. For the real matrices ($\beta=1$) Edelman has provided an efficient recurrence scheme to work out exact and explicit results for the smallest eigenvalue density which does not involve determinants or matrices. Very recently, an analogous recurrence scheme has been obtained for the complex matrices ($\beta=2$). In the present work we extend this to $\beta$-Wishart-Laguerre ensembles for the case when exponent $\alpha$ in the associated Laguerre weight function, $\lambda^\alpha e^{-\beta\lambda/2}$, is a non-negative integer, while $\beta$ is positive real. This also gives access to the smallest eigenvalue density of fixed trace $\beta$-Wishart-Laguerre ensemble, as well as moments for both cases. Moreover, comparison with earlier results for the smallest eigenvalue density in terms of certain hypergeometric function of matrix argument results in an effective way of evaluating these explicitly. Exact evaluations for large values of $n$ (the matrix dimension) and $\alpha$ also enable us to compare with Tracy-Widom density and large deviation results of Katzav and Castillo. We also use our result to obtain the density of the largest of the proper delay times which are eigenvalues of the Wigner-Smith matrix and are relevant to the problem of quantum chaotic scattering.
\end{abstract}

\section{Introduction}
The statistics of the smallest eigenvalue of the Wishart-Laguerre ensemble~\cite{Mehta2004,Forrester2010} is of considerable interest, not only from a general mathematical point of view, but also because of its connection with several interesting problems and concrete applications. A few examples of these are as follows. In principal component analysis it plays a crucial role in determining the plane of closest fit to a set of points in three-dimensional space~\cite{Gnanadesikan1997}. In the multiple channel communication it is useful in antenna selection techniques~\cite{PL2008}. It is crucial to investigating universality aspects in the spectra of QCD Dirac operators~\cite{NDW1998,DN2001}.  In compressive sensing~\cite{CT2006}, the minimal eigenvalue sets the bounds on the number of random measurements needed to fully recover a sparse signal. The smallest eigenvalue of the fixed trace variant finds application in quantum entanglement problem~\cite{MBL2008,Majumdar2011,CLZ2010,AV2011,KSA2017}. 

There are several equivalent results available for the smallest eigenvalue of Wishart-Laguerre ensembles; see, for example, Table 3 of reference~\cite{EGP2016} for the case of complex matrices ($\beta=2$). The key parameters in these representations are the dimension $n$ of the matrices and the exponent $\alpha$ in the associated Laguerre weight function $\lambda^\alpha e^{-\beta \lambda/2}$ which relates to the ``rectangularity" (see Sec.~\ref{SecBWLE}). In asymptotic analyses, some of these expressions are more suited for the hard edge limit ($n\to\infty, \alpha$ fixed) or the soft-edge limit (both $n,\alpha\to\infty$) and, therefore, can be accordingly used. Determinantal or Pfaffian-based expressions constitute prominent representations and have been explored by several authors ~\cite{Khatri1964,FH1994,Forrester1993,Forrester1994,NF1998,ZCW2009,Forrester2007,WG2013}. However, explicit evaluations by expanding the determinants or Pfaffians pose difficulties if large dimension matrices are involved. In~\cite{Edelman1989,Edelman1991} Edelman has provided a recursion based scheme for obtaining the smallest eigenvalue density for the case of real matrices ($\beta=1$) which does not involve determinants or Pfaffians and is very effective in evaluations for large-dimension cases. This approach can be used for numerical evaluation of the smallest eigenvalue density in large-dimension cases and hence enables one to verify asymptotic and universal behaviour such as those predicted by Tracy-Widom density~\cite{FS2010,TW1993,TW1994a,TW1994b} and large deviation results~\cite{KC2010}. In a recent work~\cite{KSA2017}, an analogous recurrence scheme has been derived for the complex matrices ($\beta=2$) for the smallest eigenvalue density in the unrestricted trace as well as fixed trace case. The latter result has also been used in the investigation of entanglement production~\cite{KSA2017} in the paradigmatic system of coupled kicked tops~\cite{HKS1987,Haake2010}.

In the present work we generalize the recursion scheme to the $\beta$-Wishart-Laguerre ensemble ($\beta>0$) when the parameter $\alpha$ is a non-negative integer. This enables us to work out the exact smallest eigenvalue density and corresponding moments for both unrestricted trace and fixed trace ensembles. Moreover, comparison of our smallest eigenvalue density expression with an existing result involving hypergeometric function of matrix argument~\cite{Forrester1994,Forrester2010} provides a new method for exact and explicit evaluations of these as well. We validate our analytical results with numerical simulation based on classical matrix models for $\beta=1,2,4$~\cite{Mehta2004,Forrester2010} and Dumitriu and Edelman's matrix model for general $\beta>0$~\cite{DE2002}. Moreover we perform large-dimension evaluations using our result to compare with the Tracy-Widom density~\cite{FS2010,TW1993,TW1994a,TW1994b} and large deviation results of Katzav and Castillo~\cite{KC2010}. Finally, we work out the density of the largest of proper delay times which are the eigenvalues of Wigner-Smith time-delay matrix~\cite{Eisenbud1948,Wigner1955,Smith1960,BFB1997,FS1997,BFB1999,SSS2001,Texier2016}. The latter constitute one of the key objects in the problem of quantum chaotic scattering.


\section{$\beta$-Wishart-Laguerre ensemble}
\label{SecBWLE}
\subsection{Unrestricted trace ensemble}

The joint probability density (JPD) of eigenvalues $(0<\lambda_1,...,\lambda_n< \infty)$ of the $\beta$-Wishart-Laguerre ensemble is given by
\begin{equation}
\label{JPD_WL}
P(\lambda_1,...,\lambda_n)=C_{n,\alpha,\beta}\, |\Delta_n(\{\lambda\})|^\beta \prod_{j=1}^n\lambda_j^{\alpha}\, e^{-\beta \lambda_j/2},
\end{equation}
where $\alpha>-1,\beta>0$ for convergence, and $\Delta_n(\{\lambda\})=\prod_{1\leq k<j\leq n}(\lambda_j-\lambda_k)$ is the Vandermonde determinant. The normalization factor $C_{n,\alpha,\beta}$ can be evaluated using Selberg's integral~\cite{Mehta2004,Forrester2010} as
\begin{equation}
\label{Cn}
C_{n,\alpha,\beta}=\left(\frac{\beta}{2}\right)^\gamma\,\,\prod_{j=0}^{n-1}\frac{\Gamma\big(\frac{\beta}{2}+1\big)}{\Gamma\big(\frac{\beta}{2}(j+1)+1\big)\Gamma\big(\frac{\beta}{2}j+\alpha+1\big)},
\end{equation}
with
\begin{equation}
\gamma=n\left(\alpha+\frac{\beta}{2}(n-1)+1\right).
\end{equation}
To distinguish the above from the fixed trace ensemble considered ahead, we will refer to this as a unrestricted trace (or regular) Wishart-Laguerre ensemble. 

For $\beta=1,2,4$, we have the classical matrix models for the above density in terms of real symmetric, complex-Hermitian and self-dual quaternion matrices, respectively~\cite{Mehta2004,Forrester2010}. The matrix model for these three cases can be written as
\begin{equation}
\label{betaW}
\bW=\beta^{-1}\bA\bA^\dag,
\end{equation} 
where $\bA$ is, in general, a rectangular matrix with dimensions $n\times m$, consisting of real, complex or \emph{real}-quaternion elements. It is governed by the probability density
\begin{equation}
\label{classical}
\mathcal{P}(\bA)\propto e^{-\frac{1}{2}\tr\bA\bA^\dag}
\end{equation} 
with respect to the flat measure. The operation `$\dag$' stands for transpose in case of real matrices, and conjugate-tranpose for complex matrices. For case of quaternion elements ($\beta=4$), the conjugate-transpose operation is understood over the corresponding complex matrix of double the size. We should note that for $\beta=4$,~(\ref{JPD_WL}) gives the joint density of the non-degenerate $n$ eigenvalues out of the total $2n$ produced by~(\ref{betaW}) (Kramers degeneracy). For these three special $\beta$ values the parameter $\alpha$ in~(\ref{JPD_WL}) is determined by $\beta(m-n+1)/2-1$.

Dumitriu and Edelman have provided a matrix model for the $\beta$-Wishart-Laguerre ensemble~\cite{DE2002}. The joint probability density of eigenvalues, given by~(\ref{JPD_WL}) holds for all $\beta>0$ and $\alpha>-1$ with the matrix model~(\ref{betaW}), provided $\bA$ is taken as
\begin{equation}
\label{DE}
\bA=\left(\begin{array}{ccccc}
\chi_{\beta(n-1)+2(\alpha+1)} & ~ & ~ & ~ & ~\\
\chi_{\beta(n-1)} & \chi_{\beta(n-2)+2(\alpha+1)} & ~  & ~ & ~\\
~ & \chi_{\beta(n-2)} & \chi_{\beta(n-3)+2(\alpha+1)} & ~ & ~ \\
~ & ~ & ~ & ~ & ~\\
~& ~ & \angdots{4em}{.6em} & \angdots{4em}{.6em} & ~\\
~ & ~ & ~ & \chi_\beta & \chi_{2(\alpha+1)}
\end{array}\right).
\end{equation}
Here $\chi_d$ represents a chi-distributed random variable with degrees of freedom $d$. It should be noted that the elements of $\bA$ matrix in~(\ref{DE}) above, except the diagonal and sub-diagonal ones, are all zero. 

The $\beta$-ensembles of random matrix theory with classical weights exhibit several special properties and therefore have drawn the attention of several researchers. Consequently, for these $\beta$-ensembles, there are now available a number of interesting results concerning the correlation functions, moments, fluctuation properties, the gap probabilities, and the distribution of extreme eigenvalues, etc.~\cite{DE2002,KN2004,DE2006,F2009,CMD2007,DF2006,BEMN2011,RRV2011,BN2012,DV2013,DL2014,MS2014,FRW2017}.

\subsection{Fixed trace ensemble}

It is interesting to examine the statistical behavior of a random matrix ensemble with certain constraints. Apart from a general mathematical interest, while modeling a system, these constraints can capture system dependent features~\cite{Ronsenzweig1963,Bronk1964,LP1988,ACMV1999,Mehta2004,Forrester2010,ZS2001,PPW2006,SS2015a,SS2015b}. Fixed trace ensemble is an example of such a constrained scenario and appears in, for example, quantum entanglement problem~\cite{MBL2008,Majumdar2011,CLZ2010,AV2011,KSA2017,LP1988,Page1993,ZS2001,KP2011,VPO2016,Wei2017}. 

In the case of Wishart-Laguerre ensemble, the ensemble with trace fixed at unity can be obtained using the matrices
\begin{equation}
\label{FT}
\bF=\frac{\bW}{\tr \bW},
\end{equation}
where $\bW$ is as defined in~(\ref{betaW}). 
This induces the following joint probability density for the eigenvalues (see e.g., \cite{ZS2001}):
\begin{equation}
\label{JPD_FT}
P_F(\mu_1,...,\mu_n)=K_{n,\alpha,\beta}\, |\Delta_n(\{\mu\})|^\beta \delta\left(\sum_{k=1}^n\mu_k-1\right) \prod_{j=1}^n\mu_j^{\alpha},
\end{equation}
where 
\begin{equation}
K_{n,\alpha,\beta}=\left(\frac{2}{\beta}\right)^\gamma \Gamma(\gamma)C_{n,\alpha,\beta}.
\end{equation}
It should be noted that for any other finite positive value of trace (with positive-definiteness preserved), the eigenvalue statistics can be obtained from that of unit-trace ensemble by a simple scaling.

\section{Probability density of the smallest eigenvalue: Recursion relation}
\label{SecRec}

We now present the results for the smallest eigenvalue density based on recursion scheme. 
For the unrestricted trace $\beta$-Wishart-Laguerre ensemble, the probability density of the smallest eigenvalue can be written as
\begin{equation}
\label{fW}
f(x)=c_{n,\alpha,\beta}\, e^{-\beta nx/2}x^\alpha g_{n,\alpha,\beta}(x),
\end{equation}
where the normalization factor $c_{n,\alpha,\beta}$ is
\begin{equation}
c_{n,\alpha,\beta}=\frac{n(\frac{\beta}{2})^{n\alpha+1}\Gamma\big(\frac{\beta}{2}+1\big)}{\Gamma\big(\frac{\beta n}{2}+1\big)\Gamma\big(\frac{\beta (n-1)}{2}+\alpha+1\big)}\prod_{j=0}^{n-2}\frac{\Gamma\big(\frac{\beta}{2}j+\beta+1\big)}{\Gamma\big(\frac{\beta}{2}j+\alpha+1\big)}.
\end{equation}
The scheme described below enables us to obtain $g_{n,\alpha,\beta}(x)$ if $g_{n,\alpha-1,\beta}(x)$ is known. This can be then repeated to go up to higher $\alpha$ values. Overall, the recursion involves $n$ and $\alpha$.

Consider a known $g_{n,\alpha-1,\beta}(x)$, then we apply the following recurrence scheme to arrive at $g_{n,\alpha,\beta}(x)$:
\begin{eqnarray*}
\fl&&{\bf I.~} \mathrm{ Set ~} S_0=g_{n,\alpha-1,\beta}(x), S_{-1}=0\\ 
\fl~\\
\fl&&{\bf II.~} \mathrm{ Iterate~ the~ following~ for~ }i=1 \mathrm{ ~to~ } n-1:\\
\fl&&~~~~~~S_i=\Bigg(x+\frac{2\alpha}{\beta}+n-i+1\Bigg)S_{i-1}-\frac{2x}{\beta(n-i)}\frac{d S_{i-1}}{dx}+x\,(i-1)\Bigg(1+\frac{2\alpha}{\beta(n-i)}\Bigg)S_{i-2}\\ 
\fl~\\
\fl&&{\bf III.~} \mathrm { ~Obtain~ } g_{n,\alpha,\beta}(x)=S_{n-1}
\end{eqnarray*}
Therefore, as indicated above, this recursion gives the density result corresponding to an exponent $\alpha$ if the result is known for exponent $\alpha-1$. The general integral-result for $g_{n,\alpha,\beta}(x)$ is
\begin{eqnarray}\label{gnab}
\nonumber
g_{n,\alpha,\beta}(x)=\frac{\displaystyle\int_0^\infty d\lambda_1 \cdots \int_0^\infty d\lambda_{n-1}\,|\Delta_{n-1}(\{\lambda\})|^{\beta}\prod_{j=1}^{n-1} \lambda_j^\beta e^{-\beta \lambda_j/2 }(\lambda_j+x)^\alpha}{\displaystyle\int_0^\infty d\lambda_1 \cdots \int_0^\infty d\lambda_{n-1}\,|\Delta_{n-1}(\{\lambda\})|^{\beta}\prod_{j=1}^{n-1} \lambda_j^\beta e^{-\beta \lambda_j/2}}\\
=C_{n-1,\beta,\beta}\int_0^\infty d\lambda_1 \cdots \int_0^\infty d\lambda_{n-1}\,|\Delta_{n-1}(\{\lambda\})|^{\beta}\prod_{j=1}^{n-1} \lambda_j^\beta e^{-\beta \lambda_j/2 }(\lambda_j+x)^\alpha.
\end{eqnarray}
In case $\alpha$ happens to be a non-negative integer (i.e., $\alpha=0,1,2,3,...$), then the initial case corresponding to $\alpha=0$ is quite simple and happens to be $g_{n,0,\beta}(x)=1$. Consequently, the higher integer $\alpha$ values can be obtained by repeated application of the above recurrence scheme, for any $\beta>0$. The proof of the above recursion relation relies on the properties exhibited by Selberg integrals. It is outlined in appendix A and closely follows the one provided for $\beta=1$ case by Edelman~\cite{Edelman1989,Edelman1991}. Similar recurrence schemes have been derived by Forrester for computation of correlations in the $1/r^2$ quantum many body systems~\cite{F1993}. In the context of random matrix theory, these results pertain to the Jacobi ensemble and the corresponding expressions for the Laguerre ensemble can be deduced as a limiting case. It has been also shown that these recurrences are equivalent to matrix differential equations~\cite{Forrester2010,FR2012}. Moreover, analogous recursion relations have been obtained from a viewpoint of matrix difference equations in~\cite{FI2010}.

It can be seen from the recursion scheme that for non-negative integer $\alpha$, $g_{n,\alpha,\beta}(x)$ is a polynomial of degree $\alpha(n-1)$ in $x$. Therefore, owing to the prefactor of $x^\alpha$ in~(\ref{fW}), we have
\begin{equation}\label{fW1}
f(x)=e^{-\beta nx/2}\sum_{j=\alpha}^{\alpha n}\kappa_j x^j.
\end{equation} 
Here the coefficients $\kappa_j$ are $n,\alpha,\beta$ dependent and the corresponding numerical values can be extracted easily once the recurrence has been applied. 

We show in appendix B that $\kappa_r$ can also be obtained in the following manner. Consider, for $\alpha\leq r\leq n\alpha$, the partitions of $r-\alpha$ using $n-1$ non-negative integers which are less than or equal to $\alpha$. Suppose there are $L$ such unique partitions (up to ordering):
\begin{equation}\label{partition}
\big\{\underbrace{\p_{j,1},...,\p_{j,1}}_{ s_{j,1} \mathrm{times}},\underbrace{\p_{j,2},...,\p_{j,2}}_{s_{j,2} \mathrm{times}},\ldots,\underbrace{\p_{j,l_j},...,\p_{j,l_j}}_{s_{j,l_j} \mathrm{times}}\big\};~~~j=1,...,L.
\end{equation}
It should be noted that $\p_{j,k}$ occurs with a multiplicity of $s_{j,k}$. Clearly, $\sum_{k=1}^{l_j}s_{j,k}=n-1$, and $\sum_{k=1}^{l_j}s_{j,k}\,\p_{j,k}=r-\alpha$. Then $\kappa_r$ is given by
\begin{eqnarray}
\label{coeff}
\nonumber
\kappa_r &=&(n-1)!\,c_{n,\alpha,\beta}\sum_{j=1}^L \left(\prod_{k=1}^{l_j} \frac{1}{s_{j,k}!}\binom{\alpha}{\p_{j,k}}^{s_{j,k}}\right)\\
&&\times\expval{\lambda_1^{\alpha-\p_{j,1}}\lambda_2^{\alpha-\p_{j,1}}\ldots \lambda_{s_{j,1}}^{\alpha-\p_{j,1}} \lambda_{s_{j,1}+1}^{\alpha-\p_{j,2}}\ldots \lambda_{s_{j,2}}^{\alpha-\p_{j,2}} \ldots \lambda_{n-1}^{\alpha-\p_{l_j}}}_\Lambda,
\end{eqnarray} 
where
\begin{eqnarray}
\nonumber
\expval{\mathcal{F}(\lambda_1,...,\lambda_{n-1})}_\Lambda&=&C_{n-1,\beta,\beta}\int_0^\infty d\lambda_1 \cdots \int_0^\infty d\lambda_{n-1}\,\mathcal{F}(\lambda_1,...,\lambda_{n-1})\,\\
&&\times|\Delta_{n-1}(\{\lambda\})|^{\beta}\prod_{j=1}^{n-1} \lambda_j^\beta e^{-\beta \lambda_j/2 }.
\end{eqnarray}
Therefore, we see that the calculation of $\kappa_r$ involves various moments of the eigenvalues $\lambda_1,\ldots,\lambda_{n-1}$ which, in principle, can be calculated using the theory of Selberg's integral~\cite{Mehta2004,Forrester2010,SSW2008}. However, the calculation involved can become quite cumbersome. On the other hand, the recursive method described above gives a very powerful and effective way of evaluating these coefficients, which can be then be used to write down the smallest eigenvalue density for the fixed trace case also, as shown below.

The smallest eigenvalue density for the fixed trace ensemble can be obtained from that of the unrestricted trace by applying the inverse Laplace transform~\blue{\cite{AV2011}}:
\begin{eqnarray}\label{fF}
\nonumber
f_F(x)&&=\frac{2}{\beta}\,\Gamma(\gamma)\mathcal{L}^{-1}\left[s^{1-\gamma}f\left(\frac{2sx}{\beta}\right);s;t=1\right]\\
&&=\frac{2}{\beta}\,\Gamma(\gamma)\sum_{j=\alpha}^{n\alpha}\kappa_j\frac{(1-nx)^{\gamma-j-2}}{\Gamma(\gamma-j-1)}\left(\frac{2x}{\beta}\right)^{j}\Theta(1-nx).
\end{eqnarray}
This again possesses a very simple structure and can be easily evaluated.

We now discuss how the recursive scheme can be applied for the three classical cases $\beta=1,2$ and $4$, for which we have matrix-models in terms of real-symmetric, complex-Hermitian, and self-dual quaternion matrices, respectively. Subsequently, we talk about $\beta>0$ with non-negative integer $\alpha$ where Dumitriu and Edelman's $\beta$-matrix model turns out to be useful. 
\pagebreak 
\begin{itemize}
\item
$\boldsymbol {\beta=1}$

In this case, the exponent $\alpha$ in~(\ref{JPD_WL}) is $(m-n-1)/2.$ Therefore, as rectangularity $m-n$ takes the values $0,1,2,3...$, the parameter $\alpha$ takes the values $-1/2,0,1/2,1,...$, respectively. We note that for even rectangularity, $\alpha$ assumes half-integer values, while for off rectangularity, it assumes integer values. 

Using the above recurrence scheme, the odd rectangularity cases can be handled easily. So, for $m-n=1$ (or $\alpha=0$), we have $g_{n,0,1}(x)=1$. As can be seen from~(\ref{fW}), the corresponding smallest eigenvalue density expression is quite simple in this case and is well known from earlier works~\cite{Edelman1989,Edelman1991,Forrester1993}. Applying the recurrence once on this result gives the smallest eigenvalue density for $\alpha=1$, i.e., $m-n=3$; applying it twice yields $\alpha=2$, i.e., $m-n=5$ result, and so on. 

Even rectangularity cases cannot be handled directly using the above recurrence, as in this setting the initial value $g_{n,-1/2,1}(x)$ is non-trivial. However, if we use $g_{n,-1/2,1}(x)=[n\Gamma(\frac{n+1}{2})/(\sqrt{2\pi}\,c_{n,-1/2,1})]U(\frac{n-1}{2},-\frac{1}{2};\frac{x}{2})$ based on the $\alpha=-1/2$ result of references~\cite{Edelman1989,Edelman1991,AGKWW2014,WAGKW2015}, then it is possible to apply the recursion to obtain the results for higher half-integer $\alpha$ values. Here $U(a,b;z)$ is the Tricomi's (confluent hypergeometric) function. This gives an alternative way of handling the $m-n=$ even cases, for which Edelman has provided a different recurrence scheme~\cite{Edelman1989,Edelman1991}. \\

We should also remark that explicit closed form expressions for the smallest eigenvalue density and the associated gap probability for $\beta=1$ are also known in terms of Pfaffians for this even topology scenario~\cite{AGKWW2014,WAGKW2015}. These results relate to the expectation value of the square root of the characteristic polynomial in the orthogonal ensemble~\cite{RRV2011,AGKWW2014,WAGKW2015,FN2015}.

\item
$\boldsymbol {\beta=2}$

In this case, the exponent $\alpha$ in~(\ref{JPD_WL}) is same as the rectangularity $m-n$, and hence it is very convenient to apply the above recursion~\cite{KSA2017}. The initial case here corresponds to $\alpha=0$ (or $m-n=0$), and hence $g_{n,0,2}(x)=1$. The higher rectangularity cases $m-n=1,2,3,...$ can be obtained by repetitive implementation of the recurrence relation. Equivalent representations for the smallest eigenvalue distribution in this case can be found in several earlier works, see for example ~\cite{NDW1998,DN2001,Forrester1993}. The reference~\cite{EGP2016} nicely summarizes various exact results in its Table 3.

\item
$\boldsymbol {\beta=4}$

In this case, the exponent $\alpha$ in~(\ref{JPD_WL}) is $2m-2n+1$. Therefore, the rectangularity $m-n=0,1,2,3...$, respectively, correspond to $\alpha=1,3,5,7,...$~. This case can again be handled using the recursion. We start from $\alpha=0$ (with $g_{n,0,4}(x)=1$), and apply the recurrence odd number of times to obtain the results for different rectangularities. We should remark here that the probabilty density of the smallest eigenvalue for symplectic case has been obtained in~\cite{BMW1998} in connection with the investigation of universal aspects associated with QCD Dirac operator. Morever, the scaled distribution of the extreme eigenvalues, and their asymptotics can be found for the classical random matrix ensembles in~\cite{Forrester1993} for $\beta=4$ along with $\beta=1,2$.

\item
$\boldsymbol {\beta>0}$ {\bf with non-negative integer} $\boldsymbol{\alpha} $

The smallest eigenvalue density result for $\alpha=0$ and arbitrary $\beta>0$ is well known, and was derived in reference~\cite{Forrester1993}. For this case, as noted above, $g_{n,0,\beta}(x)=1$. Therefore, we can start with $\alpha=0$ and go up to the desired value of the non-negative exponent $\alpha$. We should mention that for $\beta\neq 1,2,4$, the parameter $m$ does not have a direct interpretation as dimension of a matrix. 

\end{itemize}

Mathematica~\cite{Mathematica} codes to evaluate the above densities using the recursive scheme are given in appendix C. Tables~\ref{TbfW} and~\ref{TbfF} compile the densities for a few cases for the unrestricted ensemble and the fixed trace ensemble, respectively. We note that in the Table~\ref{TbfW}, the expressions in the second row ($\beta=1$) and the fifth row ($\beta=4$) can be obtained using the Pfaffian based results involving associated Laguerre polynomials given in~\cite{NF1998}. Similarly, the expression in the third row ($\beta=2$) can be recovered from the determinantal results derived in~\cite{FH1994}.

\begin{figure*}[ht!]
\centering
\includegraphics[width=0.9\textwidth]{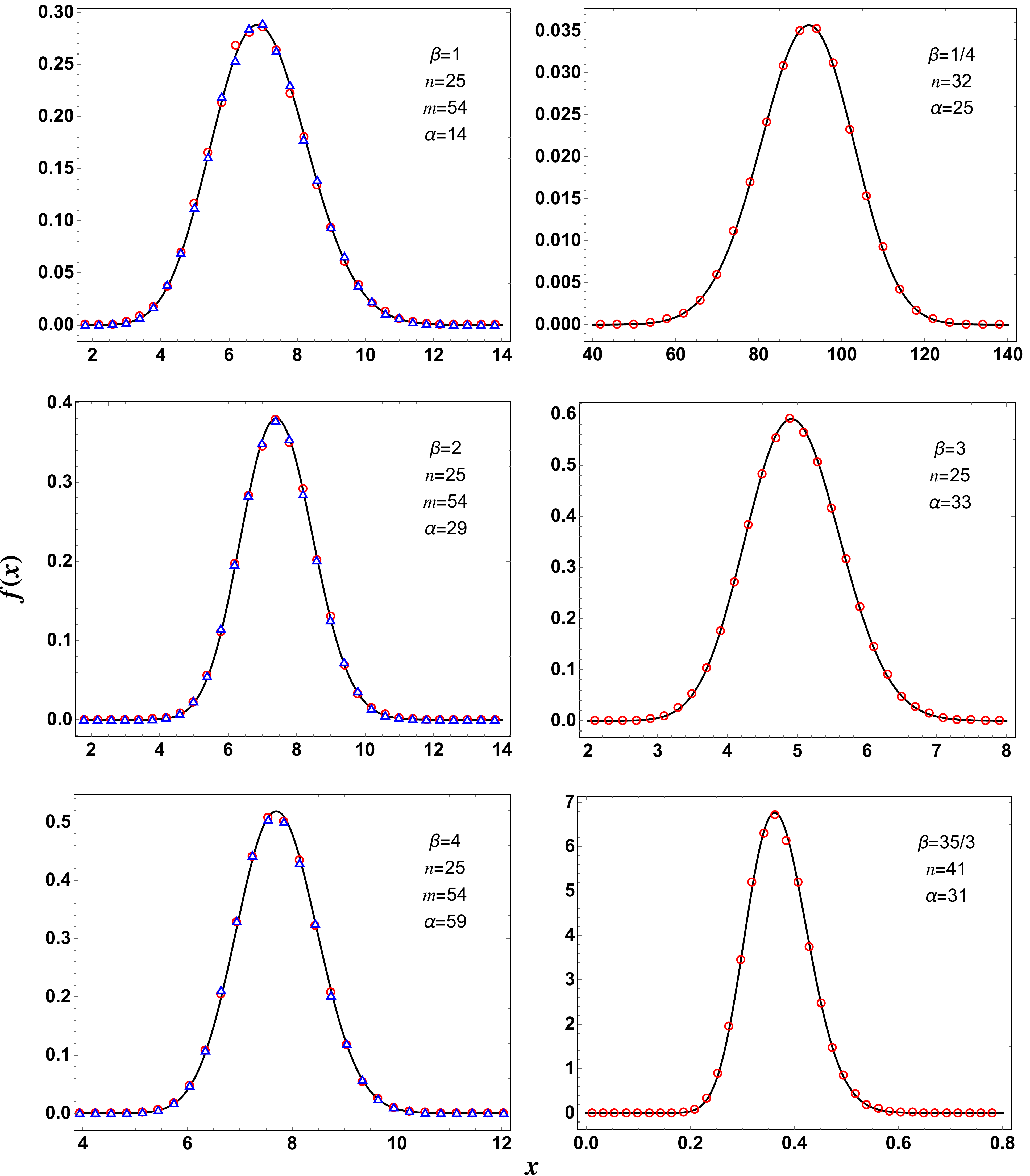}
\caption{\small Probability density of the smallest eigenvalue of unrestricted Wishart-Laguerre ensemble for several parameters values, as indicated. Solid curves (black) are based on~(\ref{fW}), circles (red) are obtained using simulation of the $\beta$-matrix model~(\ref{betaW}),~(\ref{DE}) and triangles (blue) are using simulation of classical matrix models based on~(\ref{betaW}),~(\ref{classical}).}  
 \label{FigbetaW}
\end{figure*}

\begin{figure*}[ht!]
\centering
\includegraphics[width=0.9\textwidth]{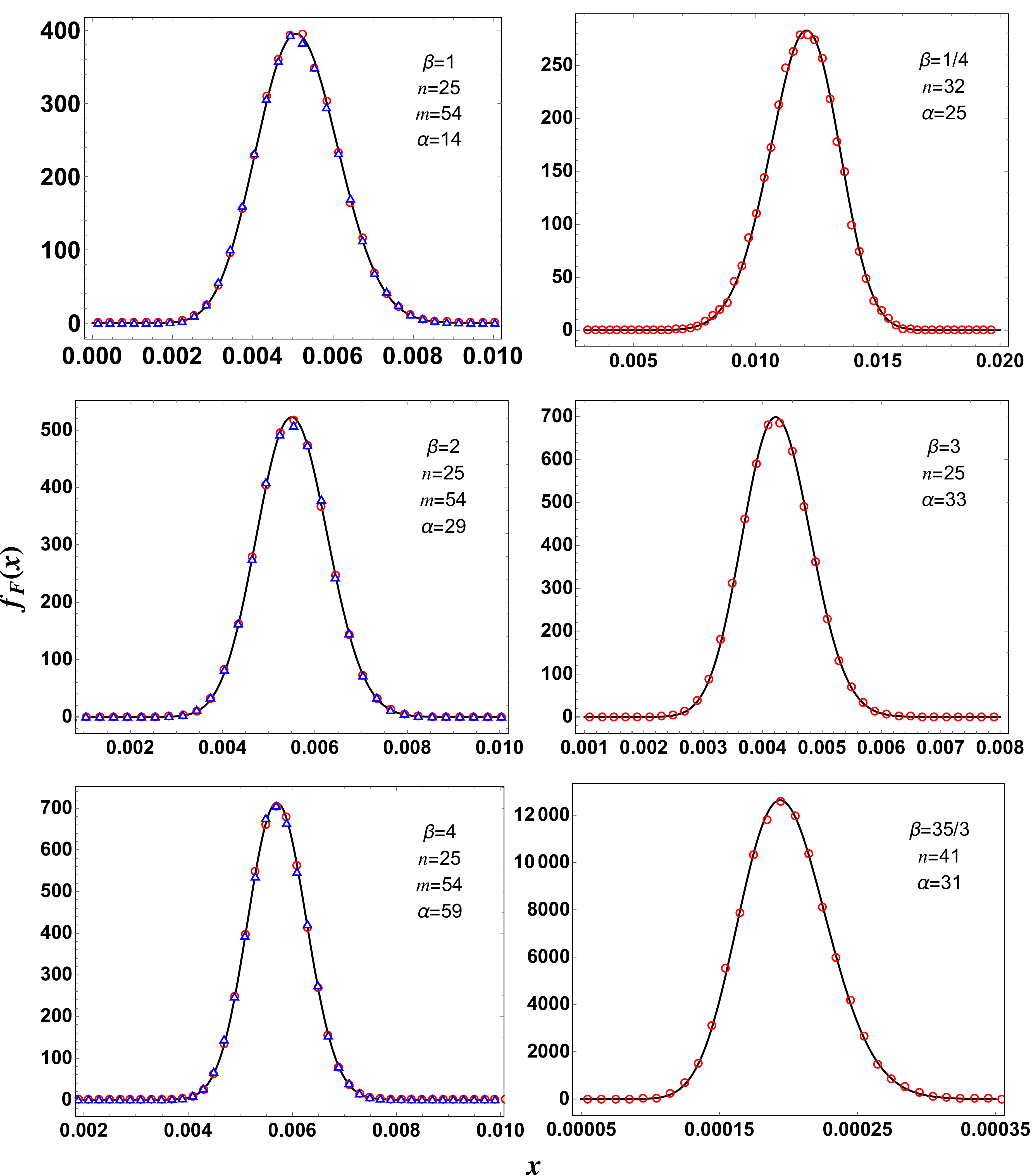}
\caption{\small Probability density of the smallest eigenvalue of fixed trace Wishart-Laguerre ensemble for several parameter values, as indicated. Solid curves (black) are based on~(\ref{fW}), circles (red) are obtained using simulation of the $\beta$-matrix model~(\ref{betaW}),~(\ref{DE}) along with~(\ref{FT}), and triangles (blue) are using simulation of classical matrix models based on~(\ref{betaW}),~(\ref{classical}) and~(\ref{FT}).}  
 \label{FigbetaF}
\end{figure*}

In Figs.~\ref{FigbetaW} and~\ref{FigbetaF} we compare the analytical result based on~(\ref{fW}) and~(\ref{fF}) with the numerically obtained densities using above mentioned matrix models with 50000 matrices. For $\beta=1,2,4$ we use the classical matrix models as well as Dumitriu and Edelman's matrix model. We can see excellent agreement in all cases.
\begin{table}[h!]
\footnotesize
\renewcommand{\arraystretch}{1.6}
\caption{\normalsize Explicit results for the smallest eigenvalue density for the unrestricted trace $\beta$-Wishart-Laguerre ensemble in a few cases. }
\centering
\begin{tabular}{|c|c| }
\hline
Parameters  & $f(x)$ \\
 \hline\hline
\makecell{$n=4$\\$\alpha=3$\\$\beta=1/2$} & \makecell{$e^{-x}x^3(x^9+72 x^8+2520 x^7+54768 x^6+804384 x^5+8297856 x^4$\\
$+60230016 x^3+300174336 x^2+958003200 x+1490227200)/217945728000$}   \\
\hline
\makecell{$n=3$\\$\alpha=4$\\$\beta=1$} & \makecell{$e^{-3x/2}x^4(x^8+40 x^7+800 x^6+10080 x^5+85680 x^4+504000 x^3$\\
$+2056320 x^2+5322240 x+6652800)/464486400$}   \\
\hline
\makecell{$n=5$\\$\alpha=2$\\$\beta=2$} & \makecell{$e^{-5x}x^2(x^8+48 x^7+960 x^6+10320 x^5+64800 x^4+241920 x^3$\\
$+524160 x^2+604800 x+302400)/17280$}   \\
\hline
\makecell{$n=3$\\$\alpha=2$\\$\beta=e$} & \makecell{$e^{-3 e x/2}\big[3 e^7 x^6+(24 e^6+36 e^7) x^5+(96 e^5+240 e^6+144 e^7) x^4$\\
$+(192 e^4+624 e^5+648 e^6+216 e^7) x^3+(192 e^3+720 e^4+960 e^5$   \\
$+540 e^6+108 e^7) x^2\big]/\big[64 (e+1) (e+2)^2 (e+4)\big]$}\\
\hline
\makecell{$n=3$\\$\alpha=3$\\$\beta=4$} & \makecell{$16 e^{-6 x} x^3 \left(4 x^6+84 x^5+735 x^4+3360 x^3+8400 x^2+11340 x+6615\right)/1575$}\\
\hline
\end{tabular}
\label{TbfW}
\qquad
\end{table}
\begin{table}[h!]
\footnotesize
\renewcommand{\arraystretch}{1.6}
\caption{\normalsize Explicit results for the smallest eigenvalue density for the fixed trace $\beta$-Wishart-Laguerre ensemble in a few cases. For compactness, we have not shown the overall factor of $\Theta(1-nx)$ in density expressions.}
\centering
\begin{tabular}{|c|c| }
\hline
Parameters  & $f_F(x)$ \\
 \hline\hline
\makecell{$n=3$\\$\alpha=4$\\$\beta=1/5$} & \makecell{$220712943321 (1 - 3 x)^{8/5} x^4(68397 x^8-122040 x^7+74044 x^6-16200 x^5$\\
$+1998 x^4-2120 x^3+1276 x^2-440 x+77)/305834375$}\\
\hline
\makecell{$n=4$\\$\alpha=3$\\$\beta=1$} & \makecell{$-36480 (1 - 4 x)^8 x^3(94976 x^9+159488 x^8-288960 x^7+197120 x^6-77728 x^5$\\
$+12768 x^4+728 x^3-112 x^2-27 x-12)$}\\
\hline
\makecell{$n=5$\\$\alpha=2$\\$\beta=2$} & \makecell{$628320  (1 - 5 x)^{23}x^2(75355 x^8-92420 x^7+29788 x^6+4676 x^5- 580 x^4$\\
$-1234 x^3+142 x^2+22 x+1)$}\\
\hline
\makecell{$n=3$\\$\alpha=2$\\$\beta=\pi$} & \makecell{$9 (3 \pi+ 2) (3 \pi+ 4) (3 \pi+ 7) (3 \pi+8 ) (1-3 x)^{3 \pi+1} x^2\big[(42+9 \pi)x^4-36x^3$\\
$+(8-6\pi)x^2-4x+(\pi+2)\big]/\big[2 (\pi+2) (\pi+4)\big]$}\\
\hline
\makecell{$n=4$\\$\alpha=3$\\$\beta=4$} & \makecell{$7238088 (1 - 4 x)^{26} x^3(3472 x^9-44528 x^8+63564 x^7-53204 x^6+23884 x^5$\\
$-2940 x^4-749 x^3+43 x^2+27 x+3)$}\\
\hline
\end{tabular}
\label{TbfF}
\qquad
\end{table}

\section{Moments of the smallest eigenvalue}
\label{SecMoments}

The simple structures of densities in~(\ref{fW1}) and~(\ref{fF}) lead to a very easy evaluation of the corresponding moments. For the unrestricted trace case, the $\eta$-th moment is given by
\begin{equation}
\label{expxn}
\expval{x^\eta}=\int_0^\infty dx\,x^\eta f(x) =\sum_{j=\alpha}^{n\alpha}\kappa_j \left(\frac{2}{\beta n}\right)^{j+\eta+1}\Gamma(j+\eta+1).
\end{equation}
Here Re$(\eta)>-\alpha-1$. Similarly, the moment in the case of the fixed trace ensemble is
\begin{eqnarray}
\label{expxfn}
\nonumber
\expval{x^\eta}_F&=&\int_0^\infty dx\,x^\eta f_F(x) =\left(\frac{\beta}{2}\right)^\eta \frac{\Gamma(\gamma)}{\Gamma(\gamma+\eta)}\expval{x^\eta}\\
&=&\frac{\Gamma(\gamma)}{\Gamma(\gamma+\eta)}\sum_{j=\alpha}^{n\alpha}\kappa_j \left(\frac{2}{\beta}\right)^{j+1}\frac{\Gamma(j+\eta+1)}{n^{j+\eta+1}}.
\end{eqnarray}
We provide Mathematica~\cite{Mathematica} codes for evaluation of the above moments in appendix C along with other codes.

The above approach has also been used in~\cite{AV2011} to obtain the moment expression for $\beta=1$ in the fixed trace case, albeit as a sum involving Beta functions. This result is equivalent to~(\ref{expxfn}) with $\beta=1$.

\section{Exact Evaluations of a class of Hypergeometric Functions with Matrix Argument}
\label{SecHyp}

In this section we show that the recurrence scheme of section~\ref{SecRec} can also be used for the exact evaluation of a class of hypergeometric functions with matrix argument.

Consider the case of  $\alpha$ being a non-negative integer. Then the integral appearing in~(\ref{gnab}), viz.
$$\int_0^\infty d\lambda_1 \cdots \int_0^\infty d\lambda_{n-1}\,|\Delta_{n-1}(\{\lambda\})|^{\beta}\prod_{j=1}^{n-1} \lambda_j^\beta e^{-\beta \lambda_j/2 }(\lambda_j+x)^\alpha,$$
can be expressed in terms of a hypergeometric function of matrix argument as~\cite{Forrester1994,Forrester2010}
$ _1F_1^{(\beta/2)}(-n+1;2\alpha/\beta+2;-x \mathds{1}_\alpha)/C_{n-1,\alpha+\beta,\beta}.$
Forrester has given the smallest eigenvalue density for $\beta$-Wishart-Laguerre ensemble in terms of this hypergeometric function~\cite{Forrester1994,Forrester2010}. If we use this hypergeometric function representation for the above multiple integeral in~(\ref{gnab}), we arrive at
\begin{equation}
\nonumber
g_{n,\alpha,\beta}(x)=\frac{C_{n-1,\beta,\beta}}{C_{n-1,\alpha+\beta,\beta}}~_1F_1^{(\beta/2)}(-n+1;2\alpha/\beta+2;-x \mathds{1}_\alpha),
\end{equation}
which, in turn, on using~(\ref{Cn}) leads to
\begin{equation}
\label{hypergeom}
\fl
_1F_1^{(\beta/2)}(-n+1;2\alpha/\beta+2;-x \mathds{1}_\alpha)=\left(\frac{\beta}{2}\right)^{\!\alpha(n-1)}\,\prod_{j=0}^{n-2} \frac{\Gamma(\beta j/2+\beta+1)}{\Gamma(\beta j/2+\alpha+\beta+1)}\cdot g_{n,\alpha,\beta}(x).
\end{equation}
Hence, this hypergeometric function can be evaluated using the recursion scheme of section \ref{SecRec}. We should note that this is different from the algorithm used by Koev and Edelman for the evaluation of such hypergeometric functions~\cite{KE2006}, which relies on the use of Jack polynomials. We note that $_1F_1^{(\beta/2)}(-n+1;2\alpha/\beta+2;-x \mathds{1}_\alpha)$ is also a polynomial of degree $\alpha(n-1)$ in $x$. In Table~\ref{1F1} we compile some explicit results for this hypergeometric function for a few sets of parameter values. Numerical evaluations at certain $x$ values have also been carried out and agree well with those obtained using the computational codes~\cite{Koev2008} based on~\cite{KE2006}
\begin{table}[h!]
\footnotesize
\renewcommand{\arraystretch}{1.6}
\caption{\normalsize Evaluation of $_1F_1^{(\beta/2)}\left(-n+1;2\alpha/\beta+2;-x \mathds{1}_\alpha\right)$ using~(\ref{hypergeom}). }
\centering
\begin{tabular}{|c|c|c|c| }
\hline
 \multirow{2}{*}{Parameters}  & \multirow{2}{*}{$_1F_1^{(\beta/2)}\left(-n+1;2\alpha/\beta+2;-x \mathds{1}_\alpha\right)$} &  \multicolumn{2}{c|}{Numerical Value}   \\ \cline{3-4}
& & $x$ & $_1F_1^{(\beta/2)}$ \\
\hline\hline
\makecell{$n=3$\\$\alpha=6$\\$\beta=1/3$} & \makecell{$x^{12}/6976704288153600+x^{11}/64599113779200$\\
$+x^{10}/1009361152800+5 x^9/107665189632$\\
$+367 x^8/215330379264+17 x^7/337296960$\\
$+11731 x^6/9612963360+1447 x^5/59339280$\\
$+2705 x^4/6781632+265 x^3/51376+49 x^2/988+6 x/19+1$}  & 10 & $22.6555$ \\
\hline
\makecell{$n=4$\\$\alpha=5$\\$\beta=1$}& \makecell{$x^{15}/83691159552000+x^{14}/929901772800$\\
$+x^{13}/20664483840+199 x^{12}/139485265920+71 x^{11}/2324754432$\\
$+1321 x^{10}/2641766400+487 x^9/75479040+x^8/14976$\\
$+67 x^7/119808+177 x^6/46592+4857 x^5/232960$\\
$+1063 x^4/11648+2705 x^3/8736+10 x^2/13+5 x/4+1$}  & 5 & $335.899$ \\
\hline
\makecell{$n=5$\\$\alpha=3$\\$\beta=2$}  & \makecell{$x^{12}/508032000+x^{11}/7056000+13 x^{10}/2822400+47 x^9/529200$\\
$+53 x^8/47040+83 x^7/8400+3091 x^6/50400+19 x^5/70$\\
$+477 x^4/560+13 x^3/7+27 x^2/10+12 x/5+1$}  & 8 & $87447.5$ \\
\hline
\makecell{$n=5$\\$\alpha=4$\\$\beta=3$}  & \makecell{$177147 x^{16}/25372857782272000+59049 x^{15}/93282565376000$\\
$+177147 x^{14}/6663040384000+911979 x^{13}/1332608076800$\\
$+160584849 x^{12}/13326080768000+3846933 x^{11}/25049024000$\\
$+333153 x^{10}/227718400+3888 x^9/366275$\\
$+118062279 x^8/1992536000+9085527 x^7/35581000$\\
$+82863 x^6/97750+42039 x^5/19550+802359 x^4/195500$\\
$+2439 x^3/425+666 x^2/119+24 x/7+1$}  & 2  & $320.040$ \\
\hline
\makecell{$n=7$\\$\alpha=3$\\$\beta=4$}  & \makecell{$64 x^{18}/488950811724375+64 x^{17}/3621857864625$\\
$+304 x^{16}/278604451125+3104 x^{15}/75983032125$\\
$+8252 x^{14}/7960127175+68356 x^{13}/3618239625$\\
$+3881 x^{12}/15181425+145856 x^{11}/55665225+3856 x^{10}/187425$\\
$+8392 x^9/67473+108502 x^8/187425+5792 x^7/2805$\\
$+65860 x^6/11781+24268 x^5/2145+4567 x^4/273$\\
$+60512 x^3/3465+256 x^2/21+36 x/7+1$}  & 1  & $72.2218$ \\
\hline
\makecell{$n=3$\\$\alpha=2$\\$\beta=5\pi$}  & \makecell{$[625 \pi ^4x^4+(1000 \pi ^3+7500 \pi ^4)x^3+(800 \pi ^2+10000 \pi ^3+30000 \pi ^4)x^2$\\
$+(320 \pi +5200 \pi ^2+27000 \pi ^3+45000 \pi ^4)x+(22500 \pi ^4+22500 \pi ^3$\\
$+64+1200 \pi +8000 \pi ^2)]/[4 (1+5 \pi ) (2+5 \pi ) (2+15 \pi ) (4+15 \pi )]$}  & 7 & $203.910$ \\
\hline
\end{tabular}
\label{1F1}
\qquad
\end{table}

\section{Evaluations for large $n$, $\alpha$}

In this section we explore the behavior of the smallest eigenvalue density for large $n$, $\alpha$ values. It is well known that the quantitative behavior of the eigenvalues is very different when either $n\to\infty$ with $\alpha$ fixed, leading to a hard edge (presence of the origin is felt), or when both $n,\alpha$ are sent to infinity, giving rise to a soft edge. In the former case, eigenvalue distributions comprise the Bessel kernel~\cite{TW1994b,FH1994,NDW1998,DN2001}. On the other hand, in the latter case one encounters the Airy kernel which then leads to the celebrated Tracy-Widom (TW) distribution~\cite{TW1993,TW1994a}. In~\cite{BF2003} the transition regime between the Bessel and Airy densities has also been investigated. We should emphasize that the expansion of determinant or Pfaffian expressions become particularly impractical while dealing with the soft-edge scenario when both $n$ and $\alpha$ are large because the size of matrices involved depends either on $n$ or $\alpha$. In such a situation the recursion based scheme turns out to be quite useful, as will be seen below.

\subsection{Comparison with Tracy-Widom density}

Feldheim and Sodin~\cite{FS2010} have proved for $\beta=1$ and 2 cases that, in the limit $m\rightarrow \infty$, $n\rightarrow \infty$ with $n/m$ bounded away from 1 (the soft-edge scenario), the shifted and scaled smallest eigenvalue, $(\lambda_{\rm min}-\nu)/\sigma$, leads to the Tracy-Widom density~\cite{TW1993,TW1994a}. Here $\nu=(n^{1/2}-m^{1/2})^2$ and $\sigma=(n^{1/2}-m^{1/2})(n^{-1/2}-m^{-1/2})^{1/3}<0$. In~\cite{Ma2012} some modifications in these parameters have been suggested for a better agreement with Tracy-Widom density. However, we stick to the parameters defined in~\cite{FS2010}. It should be noted that the square case ($m=n$) constitutes the extreme hard-edge scenario, for which the scaled density is an exponential. In reference~\cite{CLZ2010} it has been shown that after proper scaling, the smallest eigenvalue density for fixed trace scenario behaves identical to that of the unrestricted case. Therefore, in this case also the Tracy-Widom density is expected in the asymptotic limit.

\begin{figure*}[ht!]
\centering
\includegraphics[width=0.85\textwidth]{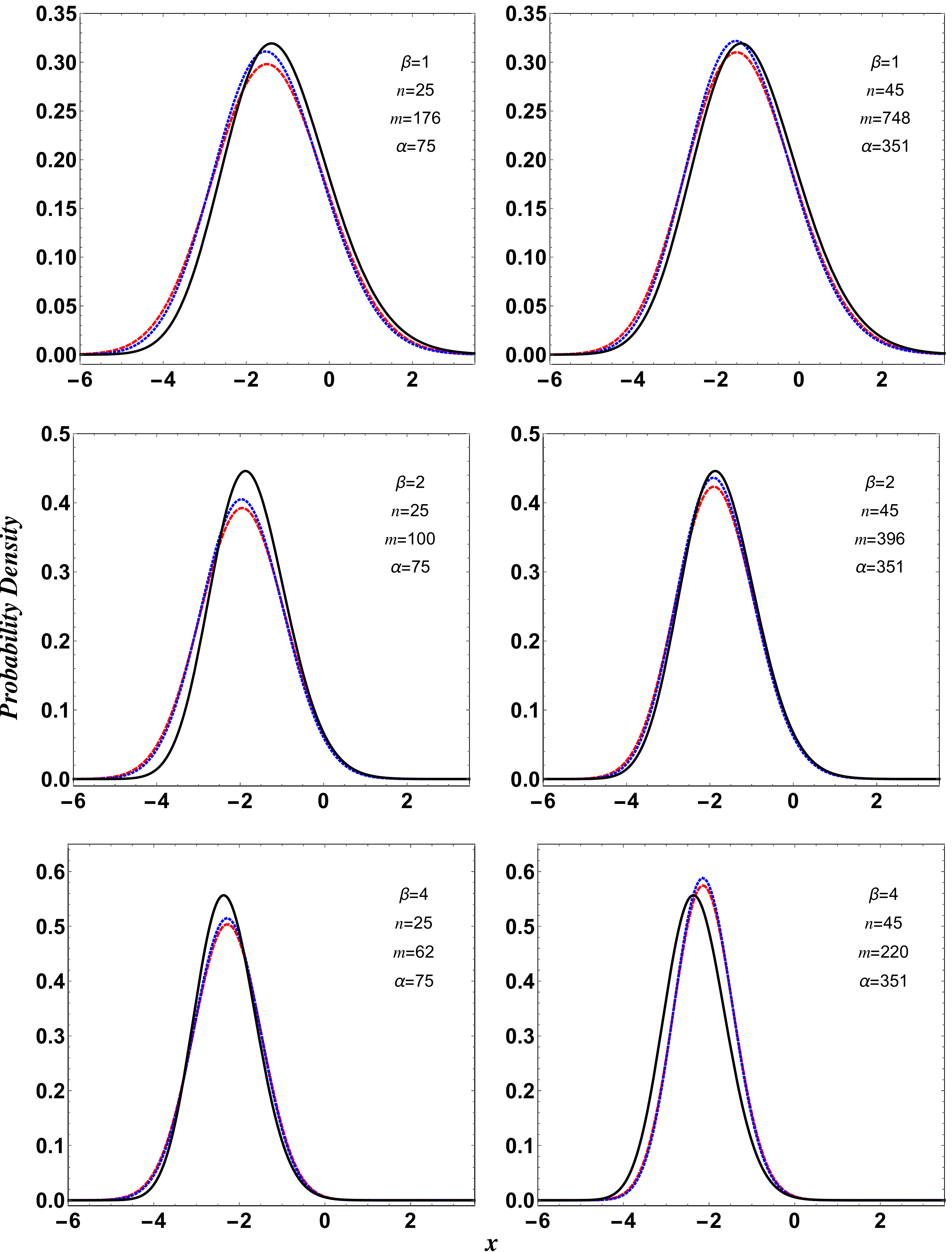}
\caption{\small Comparison of density of the transformed smallest eigenvalue with Tracy Widom density for $\beta=1,2,4$ for some $n,m$ (equivalently $\alpha$) values, as indicated. The dashed (red) and dotted (blue) curves correspond to the unrestricted trace and fixed trace cases, respectively. The solid (black) lines show the Tracy-Widom density.}  
 \label{FigTW}
\end{figure*}

The recursion scheme given in section~\ref{SecRec} enables us to work out the exact results for the smallest eigenvalue density for large values of $n$ and $\alpha$ and hence to explore the soft-edge limit numerically. In view of the scaling and shift indicated above, $-\sigma f(\sigma x+\nu)$ and $-(\sigma/mn) f_F((\sigma x+\nu)/mn)$ should coincide with the Tracy-Widom density of the corresponding symmetry class; see for example~\cite{KSA2017} for $\beta=2$. The additional scaling of $mn$ in the fixed trace case has to do with the result that average trace of unrestricted Wishart-Laguerre ensemble defined by~(\ref{JPD_WL}) is $mn$ for $\beta=1,2,4$~\footnote{More generally, it is $2\gamma/\beta$ for the $\beta$-ensemble.}. While, the proof concerning the Tracy-Widom behavior has been provided only for $\beta=1,2$ in~\cite{FS2010}, we also consider $\beta=4$ and compare with the large $n,\alpha$ evaluations of the exact density. The results are shown in Fig.~\ref{FigTW}. For $\beta=1,2$ we see that the exact density curves approach the Tracy-Widom curves from left and agreement becomes better with increasing $n,\alpha$. The convergence is, however, rather slow and can be attributed to the fact that at the soft-edge the convergence goes typically as $n^{-1/6}$.

\subsection{Comparison with large deviation results}

The Tracy-Widom density discussed in the preceding subsection describes only the typical fluctuations of the smallest eigenvalue. For large atypical fluctuations one has to work out the large deviation results~\cite{MS2014}. This has been done for the Wishart-Laguerre ensemble by Katzav and Castillo in~\cite{KC2010}. We compare in this section these large deviation results with our exact results based on the recursion relation.

\begin{figure*}[ht!]
\centering
\includegraphics[width=0.85\textwidth]{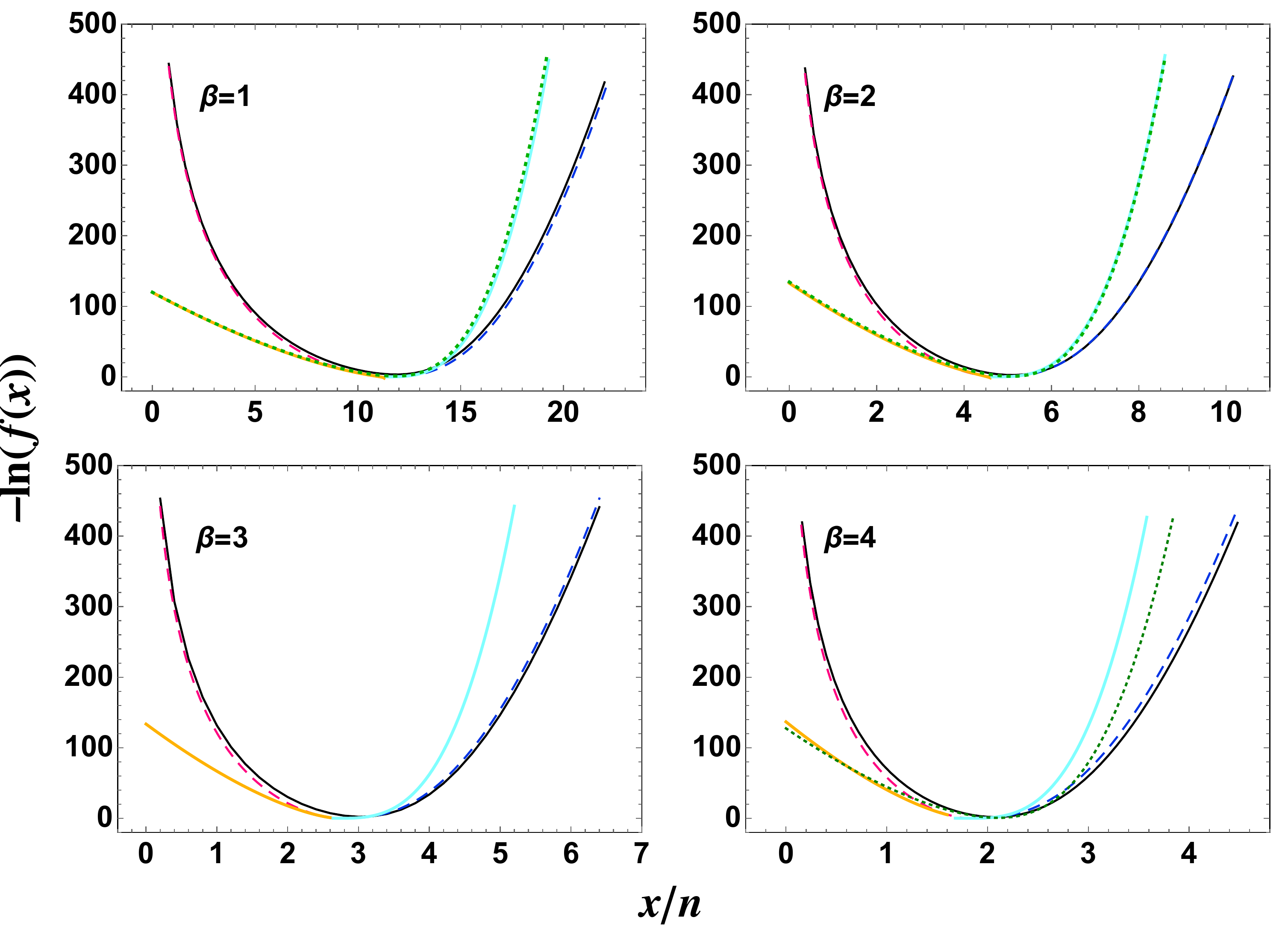}
\caption{\small Comparison of exact results with large deviation predictions for $n=25$ and $\alpha=225$. The solid (black) curves are exact results based on the recursion relation. The dashed lines are large deviation results with the curves on left (red) and right (blue) determined by the left and right rate function, respectively. For $\beta=1,2,4$ we also show the Tracy-Widom density by the dotted (green) curve. Additionally, the solid cyan and orange curves shows the prediction for the tail behavior of Tracy-Widom density. It should be noted that due to the transformation $(\lambda_{\rm min}-\nu)/\sigma$, with negative $\sigma$, the cyan curve corresponds to the left tail and the orange to the right tail.}
 \label{FigLD}
\end{figure*}

Following the reference~\cite{KC2010}, we define the parameters
\begin{equation}
\nonumber
A=\frac{2(\alpha+1)-\beta}{\beta n},
\zeta_{\pm}=(1\pm\sqrt{1+A})^2,
\Delta_{-}=\zeta_{+}-\zeta_{-}=4\sqrt{1+A},
\end{equation}
and the quantities
\begin{eqnarray}
\nonumber
\fl
&&P\equiv P(\zeta)=-\zeta-2(A+2),~ Q\equiv Q(\zeta)=2A\sqrt{\zeta},~~
B\equiv B(\zeta)=-\left(\frac{P^3}{27}+\frac{Q^2}{4}\right),\\
\nonumber
\fl
&&R\equiv R(\zeta)=\sqrt{-\frac{P^3}{27}},~~
\theta\equiv \theta(\zeta)=\tan^{-1}\left(\frac{2\sqrt{B}}{Q}\right),
~~W\equiv W(\zeta)=\frac{2P}{3R^{1/3}}\cos\left(\frac{\theta+2\pi}{3}\right),\\
\nonumber
\fl
&&U\equiv U(\zeta)=W^2,~~\Delta\equiv \Delta(\zeta)=U-\zeta,\\
\nonumber
\fl
&&S(\zeta)=\frac{U+\zeta}{2}-\frac{\Delta^2}{32}-\ln\left(\frac{\Delta}{4}\right)+\frac{A}{4}(W-\sqrt{\zeta})^2+\frac{A^2}{4}\ln(\zeta U)-A(A+2)\ln\left(\frac{W+\sqrt{\zeta}}{2}\right).
\end{eqnarray}
Then, the asymptotic probability density is given by
\begin{eqnarray}
f(x)\sim \cases{
\exp\left(-\beta n \phi_{-}\left(\frac{n\zeta_{-}-x}{n}\right)\right), & $x\in[0,n\zeta_-]$,\\ 
\exp\left(-\beta n^2 \phi_{+}\left(\frac{x-n\zeta_{-}}{n}\right)\right), & $x\in[n\zeta_-,\infty]$,}
\end{eqnarray}
where $\phi_{-}$ and $\phi_{+}$ are left and right rate functions, given respectively by
\begin{eqnarray}
\fl
\nonumber
\phi_{-}(z)=-\frac{A}{2}\ln\left(1-\frac{z}{\zeta_{-}}\right)-\frac{\sqrt{z}\sqrt{z+\Delta_{-}}}{2}+2\ln\left(\frac{\sqrt{z+\Delta_{-}}-\sqrt{z}}{\sqrt{\Delta_{-}}}\right)\\
+A\ln\left(1+2\sqrt{\frac{z}{\zeta_{-}}} \frac{\sqrt{z+\Delta_{-}}-\sqrt{z}}{\Delta_{-}}\right).
\end{eqnarray}
\begin{equation}
\fl
\phi_{+}(z)=\frac{1}{2}\left[S(z+\zeta_{-})-S(\zeta_{-})\right].
\end{equation}
In Fig.~\ref{FigLD} we compare these results for $n=25$, $\alpha=225$ and $\beta$ values of $1,2,3,4$. We can see good agreement in all cases, and thereby corroborate the analytical predictions of Katzav and Castillo~\cite{KC2010} for the soft-edge scenario. We should emphasize that in~\cite{KC2010} the large deviation predictions could be compared with the exact analytical result only for $\beta=1$ using Edelman's recursion scheme~\cite{Edelman1989,Edelman1991}. In this figure we also show the the Tracy-Widom density curves along with the plots for its tail behavior~\cite{BBD2008,RRV2011,BEMN2011,BN2012,DV2013,MS2014}:
\begin{eqnarray}
f_\mathrm{TW}(x)\sim \cases{ \exp\left(-\frac{\beta}{24}|x|^3\right), & $ x\to-\infty$,\\
\exp\left(-\frac{2\beta}{3}x^{3/2}\right),  & $x\to+\infty$.}
\end{eqnarray}
We note that the references~\cite{DV2013,BBD2008,BEMN2011,BN2012,MS2014} also give higher order corrections to the above result. We find that, in the scale considered in the Fig.~\ref{FigLD}, higher order terms do not lead to any noticeable difference in $\beta=1,2$ cases. On the other hand, for $\beta=4$ there is some improvement in agreement with the Tracy-Widom curve and apparently one would need many terms in the tail asymptotic expansion for a good agreement, especially for the left tail.

\section{Density of the largest of proper delay times}

In this section we use our result for the smallest eigenvalue density of Wishart-Laguerre ensemble to obtain the density of the largest of proper delay times which are eigenvalues of the Wigner-Smith matrix~\cite{Eisenbud1948,Wigner1955,Smith1960}. In other words, we obtain the density of the largest eigenvalue of the Wigner-Smith matrix. This problem relates  to the notion of time delay in a quantum mechanical scattering, as put forward by Eisenbud~\cite{Eisenbud1948}, Wigner~\cite{Wigner1955} and Smith~\cite{Smith1960}. Proper delay times are very relevant to the quantum chaotic scattering problem and relate to several transport observables of interest in systems such as microwave resonators and quantum dots~\cite{BFB1997,BFB1999,BLFBB1997,SBB2001,SSS2001,MBB2014,SMB2015,Texier2016}. Consequently, considerable amount of work has been done in exploring the statistics of proper delay times and related quantities~\cite{SSS2001,MZ2011,MZ2012,MZ2013,TM2013,MBB2014,KSS2014,SMB2015,Cunden2015,CMSV2016a,CMSV2016b}.

Within the \emph{Heidelberg approach}~\cite{MW1969,VWZ1985,FS1997,KNSGDORS2013,NKSG2014} the $n\times n$ scattering matrix $\bS$ is expressed in terms of an $M\times M$ random Hermitian matrix $\bH$ and an $M\times n$ coupling matrix $\bV$,
\begin{equation}
\bS = \mathds{1}_n-2\pi i \bV^\dag(E\mathds{1}_M-\bH + i\pi \bV\bV^\dag)^{-1}\bV.
\end{equation}
The matrix $\bH$ is typically modeled using the Gaussian ensemble of random matrices,
\begin{equation}
P(\bH)\propto \exp\left(-\frac{\beta M}{4v^2}\tr \bH^2\right).
\end{equation}
Here $\beta$ (=1, 2 or 4) physically has to with the time-reversal and rotation symmetries of the system~\cite{Mehta2004,Forrester2010}. The parameter $v$ fixes the energy scale. The density of states for large $M$ is given by~\cite{Mehta2004,Forrester2010} 
\begin{equation}
R_1(E)=\frac{M}{2\pi v^2}\sqrt{4v^2-E^2},
\end{equation}
and decides the mean level spacing as $\Delta=1/R_1(E)$. The latter behaves as $\pi v/M$ around center of the spectrum.
For the case of ballistic point contacts (`ideal leads') we may take $\bV_{j,k}=\delta_{j,k}(M\Delta)^{1/2}/\pi$ considering suitable basis transformations on $\bS$ and $\bH$.
Proper delay times $\tau_1,...,\tau_n$ are the eigenvalues of Wigner-Smith time-delay matrix~\cite{Smith1960,BFB1997,FS1997,BFB1999,SSS2001,Texier2016} 
\begin{equation}
\label{Qmat}
\bQ=-i\hbar \bS^{-1}\frac{\partial \bS}{\partial E}.
\end{equation}
We note that $\bS$ and its energy derivative $\partial \bS/\partial E$ can be written with the aid of Wigner $\bK$-matrix,
\begin{equation}
\bK=\pi \bV^\dag (E\mathds{1}_M-\bH)^{-1}\bV,
\end{equation}
as
\begin{equation}
\fl
\bS=-\mathds{1}_n+2(\mathds{1}_n+i\bK)^{-1},~~~\frac{\partial \bS}{\partial E}=2\pi i(\mathds{1}_M+i\bK)^{-1}\bV^\dag(E\mathds{1}_M-\bH)^{-2}\bV(\mathds{1}_M+i\bK)^{-1}.
\end{equation}
The symmetrized matrix $\widetilde{\bQ}=-i\hbar \bS^{-1/2}(\partial \bS/\partial E)\bS^{-1/2}$ share the same eigenvalues as $\bQ$, and it is known for the case of `ideal leads' the distribution of proper delay rates $\upsilon_j=1/\tau_j$ is governed by the Wishart-Laguerre density given by~(\ref{JPD_WL}) with $\alpha\rightarrow \beta n/2$ and $\lambda_j\rightarrow \tau_\mathrm{H} \upsilon_j$~\cite{BFB1997,BFB1999}. Here $\tau_\mathrm{H}=2\pi \hbar/\Delta$ is the Heisenberg time. For $\beta=4$ we have the occurrence of Kramers degeneracy and~(\ref{JPD_WL}) describes half of the spectrum which consists of the nondegenerate eigenvalues. 

As $\tau_j=1/\upsilon_j$, it is clear that the smallest $\upsilon$ will correspond to the largest $\tau$. Therefore, the smallest eigenvalue density for Wishart-Laguerre ensemble can be used to deduce the density of the largest eigenvalue of the Wigner-Smith matrix, i.e., largest of the proper delay times. We have
\begin{equation}
\label{ftau}
\hat{f}_{\tau_{\max}}(x)=\frac{\tau_\mathrm{H}}{x^2}f\left(\frac{\tau_\mathrm{H}}{x}\right),
\end{equation}
where $f(x)$ is given by~(\ref{fW}) with $\alpha=\beta n/2$.

\begin{figure*}[ht!]
\centering
\includegraphics[width=1\textwidth]{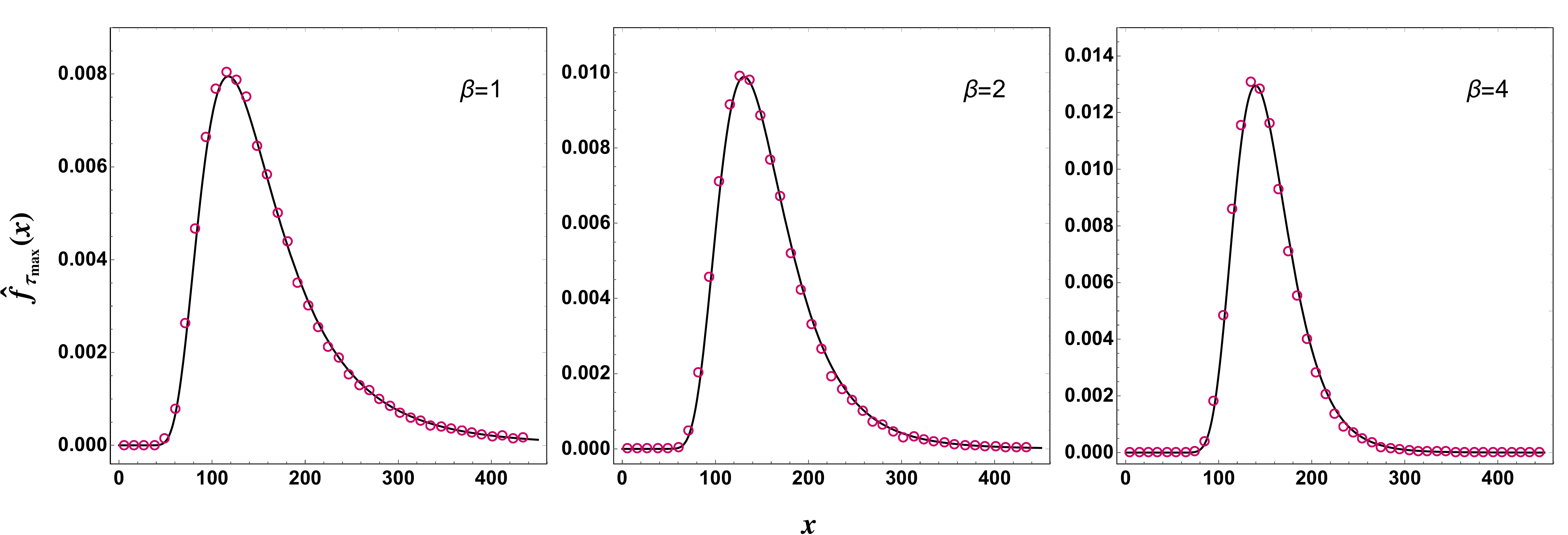}
\caption{\small Probability density of the largest of the delay times eigenvalue for $\beta=1,2,4$ and parameters indicated. Solid (black) curves are based on~(\ref{ftau}), circles (red) are obtained using simulation using~(\ref{Qmat}).}  
 \label{DT}
\end{figure*}

We verify the above by numerically simulating the $\bQ$ matrices for the three symmetry classes, i.e., $\beta=1,2,4$, obtaining its largest eigenvalue density and then comparing with~(\ref{ftau}). The parameters used for the simulation are $M=200$, $n=8$, $v=1$, $\hbar=1$, $E=0$ ({\it center of the semicircle}) with an ensemble consisting of 50000 matrices. We find excellent agreement in all cases, as can be seen in the Fig.~\ref{DT}.

We should remark that in~\cite{MBB2014} the statistics of proper time delay has also been generalized to the symmetry classes introduced by Altland and Zirnbauer~\cite{AZ1996,Zirnbauer1996}, and is again related to the Wishart-Laguerre ensemble. Consequently, our result can be extended to these as well.

\section{Summary and Conclusion}
We worked out a recurrence scheme for the evaluation of exact and explicit expressions for the smallest eigenvalue density of $\beta$-Wishart-Laguerre ensemble when the weight function $\lambda^\alpha e^{-\beta\lambda/2}$ has the exponent $\alpha$ as a non-negative integer. We considered both the unrestricted and restricted trace variants. In the process we also found a way to evaluate a class of hypergeometric function of matrix argument. Large $n,\alpha$ evaluations of the smallest eigenvalue density enabled us to directly verify the Tracy-Widom law and large-deviation asymptotic results. Finally, we used the connection of proper delay times with the Wishart-Laguerre ensemble to work out the density of the largest of the delay times which happens to be the largest eigenvalue of the Wigner-Smith matrix.

\ack{The author is grateful to Prof. Katzav for fruitful correspondences. He also thanks the anonymous reviewer whose comments helped improve the manuscript. This work has been supported by the grant EMR/2016/000823 provided by SERB, DST, Government of India.}

\appendix

\section{Proof of recurrence scheme}
\label{AppRec}

The proof of the recurrence scheme for general $\beta$ is similar to the one given by Edelman for $\beta=1$ case~\cite{Edelman1989,Edelman1991}. It was also extended recently for $\beta=2$ case~\cite{KSA2017}.
The smallest eigenvalue density can be obtained from the joint probability density of eigenvalues as
\begin{equation}
\label{fWdef}
f(x)=n\int_x^\infty d\lambda_2\cdots \int_x^\infty d\lambda_n \,P(x,\lambda_2,...,\lambda_n).
\end{equation}
We now consider $\lambda_i\rightarrow \lambda_i+x$, followed by shift of indices of the integration variables as $\lambda_i\rightarrow \lambda_{i-1}$. This gives
\begin{eqnarray}
\nonumber
\fl
f(x)=nC_{n,\alpha,\beta}\, x^\alpha e^{-\beta n x/2} &&\int_0^\infty d\lambda_1\cdots \int_0^\infty d\lambda_{n-1}\prod_{1\leq k<j\leq n-1}|\lambda_j-\lambda_k|^\beta\\
&&~~~~~~~~~~~~~~~~\times\prod_{i=1}^{n-1}\lambda_i^\beta(\lambda_i+x)^\alpha e^{-\beta\lambda_i/2}.
\end{eqnarray}
We now introduce the measure $d\Omega_i=\lambda_i^\beta \,e^{-\beta\lambda_i/2}\,d\lambda_i$ and write the above as
\begin{eqnarray}
\fl
f(x)=nC_{n,\alpha,\beta}\, x^\alpha e^{-\beta n x/2} \int_0^\infty d\Omega_1\cdots \int_0^\infty d\Omega_{n-1}|\Delta_{n-1}(\{\lambda\})|^\beta \prod_{i=1}^{n-1}(\lambda_i+x)^\alpha.
\end{eqnarray}
Following~\cite{Edelman1989,Edelman1991}, we now define
\begin{equation}
\label{def1}
I_{i,j}^\alpha=\int_0^\infty d\Omega_1\cdots\int_0^\infty d\Omega_{n-1}\, |\Delta_{n-1}(\{\lambda\})|^\beta u(x),
\end{equation} 
where 
\begin{eqnarray}
\nonumber
&&u(x)\equiv \underbrace{(\lambda_1+x)^\alpha\cdots(\lambda_i+x)^\alpha}_{i \mathrm{~ terms~}}\underbrace{(\lambda_{i+1}+x)^{\alpha-1}\cdots(\lambda_{i+j}+x)^{\alpha-1}}_{j \mathrm{ ~terms~}}\\
&&\times \underbrace{(\lambda_{i+j+1}+x)^{\alpha-2}\cdots(\lambda_n+x)^{\alpha-2}}_{n-i-j-1 \mathrm{~ terms~}}.
\end{eqnarray}
We also consider the operator
\begin{eqnarray}
\label{def2}
I_{i,j}^\alpha[v]=\int_0^\infty d\Omega_1\cdots\int_0^\infty d\Omega_{n-1} \,|\Delta_{n-1}(\{\lambda\})|^\beta u(x)\,v. 
\end{eqnarray}
Using the above, the smallest eigenvalue density can be written using~(\ref{def1}) as
\begin{equation}
\label{sev}
f(x)=nC_{n,\alpha,\beta} x^\alpha e^{-\beta nx/2}I_{n-1,0}^\alpha.
\end{equation}
Moreover, Lemma 4.2 of~\cite{Edelman1989} (or, Lemma 4.1 of~\cite{Edelman1991}) holds:
\begin{eqnarray}
\label{op_result}
I_{i,j}^\alpha[\lambda_k]=\cases{ I_{i+1,j-1}^\alpha-x I_{i,j}^\alpha & if   $i<k\leq i+j$,\\
                                  I_{i,j+1}^\alpha-x I_{i,j}^\alpha &  if   $i+j < k< n$.
                                  }
\end{eqnarray}
The above result is obtained by writing $\lambda_k$ as $(\lambda_k+x)-x$ and then using the operator defined in~(\ref{def2}). 
Now, if the terms $(\lambda_k+x)$ and $(\lambda_l+x)$ share the same exponent in the integrals (i.e., both $k$ and $l$ fall within one of the closed intervals $[1,i],[i+1,i+j]$, or $[i+j+1,n-1]$), then
\begin{eqnarray}
\label{eqa}
I_{i,j}^\alpha\left[\frac{\lambda_k\lambda_l}{\lambda_k-\lambda_l}\right]=0,\\
\label{eqb}
I_{i,j}^\alpha\left[\frac{\lambda_k}{\lambda_k-\lambda_l}\right]=\frac{1}{2}I_{i,j}^\alpha,\\
\label{eqc}
I_{i,j}^\alpha\left[\frac{\lambda_k^2}{\lambda_k-\lambda_l}\right]=I_{i,j}^{\alpha}[\lambda_k].
\end{eqnarray}
Equation~(\ref{eqa}) is a consequence of the asymmetry in $\lambda_k$ and $\lambda_l$, while ~(\ref{eqb}) is obtained using the identity $\lambda_k/(\lambda_k-\lambda_l)+\lambda_l/(\lambda_l-\lambda_k)=1$ and employing symmetry. Equation~(\ref{eqc}) follows with the aid of the identity $\lambda_k^2/(\lambda_k-\lambda_l)=\lambda_k+\lambda_k\lambda_l/(\lambda_k-\lambda_l)$ and (\ref{eqa}). 

The generalization of the Lemma 4.3 of~\cite{Edelman1989} (or Lemma 4.2~\cite{Edelman1991}) happens to be
\begin{eqnarray}
\fl
&&I_{i,j}^\alpha=\left(x+\frac{2\alpha}{\beta}+j+2k+2\right)\! I_{i-1,j+1}^\alpha-x\left(k+\frac{2(\alpha-1)}{\beta}\right)\! I_{i-1,j}^\alpha+(i-1)x I_{i-2,j+2}^\alpha \label{Rec1}\\
\fl
&&I_{0,j}^\alpha=I_{j,n-j-1}^{\alpha-1},\label{Rec2}
\end{eqnarray}
with $k=n-i-j-1$. The definition~(\ref{def1}) readily yields the second equation above, (\ref{Rec2}). The first equation of this set, (\ref{Rec1}), is derived using 
\begin{equation}
I_{i,j}^\alpha=x I_{i-1,j+1}^\alpha+I_{i-1,j+1}^\alpha[\lambda_i],
\end{equation}
which is a consequence of~(\ref{op_result}). For the case of general $\beta$, calculation of $I_{i-1,j+1}^\alpha[\lambda_i]$ involves the follwing result:
\begin{eqnarray}
\nonumber
\int_0^\infty (\lambda_i+x)^{\alpha-1}\prod_{i<l}|\lambda_l-\lambda_i|^\beta\,\lambda_i^{\beta+1}\,e^{-\beta\lambda_i/2}\,d\lambda_i\\
=\frac{2}{\beta}\int_0^\infty  \frac{d}{d\lambda_i} \bigg[(\lambda_i+x)^{\alpha-1}\prod_{i<l}|\lambda_l-\lambda_i|^\beta \,\lambda_i^{\beta+1}\bigg]e^{-\beta\lambda_i/2}\,d\lambda_i.
\end{eqnarray}
Next, using the result 
\begin{equation}
\frac{dI_{i-1,j+1}^\alpha}{dx}=(i-1)\alpha I_{i-2,j+2}^\alpha+(j+1)(\alpha-1)I_{i-1,j}^\alpha
\end{equation}
for $i+j=n-1$, we obtain (see Lemma 4.4,~\cite{Edelman1989}, or Lemma 4.3~\cite{Edelman1991})
\begin{eqnarray}
\label{Rec3}
\nonumber
I_{i,j}^\alpha=\left(x+\frac{2\alpha}{\beta}+j+2\right)I_{i-1,j+1}^\alpha-\frac{2x}{\beta(j+1)}\frac{d}{dx}I_{i-1,j+1}^\alpha\\
~~~~~~~~~~~~~~~~~~+x (i-1)\left(1+\frac{2\alpha}{\beta(j+1)}\right)I_{i-2,j+2}^\alpha.
\end{eqnarray}
For $j=n-i-1$ this yields
\begin{eqnarray}
\label{Rec4}
\nonumber
I_{i,n-i-1}^\alpha=\left(x+\frac{2\alpha}{\beta}+n-i+1\right)I_{i-1,n-i}^\alpha-\frac{2x}{\beta(n-i)}\frac{d}{dx}I_{i-1,n-i}^\alpha\\
~~~~~~~~~~~~~~~~~~+x (i-1)\left(1+\frac{2\alpha}{\beta(n-i)}\right)I_{i-2,n-i+1}^\alpha.
\end{eqnarray}
We now consider $I_{0,n-1}^\alpha$, which is same as $I_{n-1,0}^{\alpha-1}$ in view of~(\ref{Rec2}), and use~(\ref{Rec4}) repeatedly for $i=1$ to $n-1$ to arrive at $I_{n-1,0}^\alpha$. Therefore, we note that, starting from $I_{n-1,0}^{\alpha-1}$ we can arrive at $I_{n-1,0}^\alpha$, which is the term needed to obtain the smallest eigenvalue density expression~(\ref{sev}) explicitly. This is essentially what we implement in the recursion involving $S_i:= I_{i,n-i-1}^\alpha/I_{n-1,0}^0$ for $g_{n,\alpha,\beta}(x)$ in~(\ref{fW}). We also observe that $I_{n-1,0}^0=1/C_{n-1,\beta,\beta}$, which gives the constant $c_{n,\alpha,\beta}$ of~(\ref{fW}) as $nC_{n,\alpha,\beta}/C_{n-1,\beta,\beta}$.

\section{Proof of equation~(\ref{coeff})}
\label{AppCoeff}

For a non-negative integer $\alpha$, using the Binomial theorem, we have
\begin{equation}
(\lambda_j+x)^\alpha=\sum_{k=0}^{\alpha}\binom{\alpha}{k}  \lambda_j^{\alpha-k} x^k.
\end{equation}
We use this within the integral in~(\ref{gnab}). Now, since~(\ref{fW}) already contain a factor $x^\alpha$, the coefficient of $x^r$ in this equation and hence in (\ref{fW1}) is decided by the coefficient of $x^{r-\alpha}$ in $\prod_{j=1}^{n-1}\sum_{k=0}^{\alpha}\binom{\alpha}{k} \lambda_j^{\alpha-k} x^k$ which, when expanded, appears as
\begin{eqnarray*}
&\left(\binom{\alpha}{0}\lambda_1^\alpha+\binom{\alpha}{1}\lambda_1^{\alpha-1}x+\binom{\alpha}{2}\lambda_1^{\alpha-2}x^2+\cdots+\binom{\alpha}{\alpha}x^\alpha\right)\\
\times &\left(\binom{\alpha}{0}\lambda_2^\alpha+\binom{\alpha}{1}\lambda_2^{\alpha-1}x+\binom{\alpha}{2}\lambda_2^{\alpha-2}x^2+\cdots+\binom{\alpha}{\alpha}x^\alpha\right)\\
&~~~~~~~~~~~~~~~~~~~~~~~~\vdots~~~~~~~~~~~~~~~~~~~~~~~~\vdots\\
\times &\left(\binom{\alpha}{0}\lambda_{n-1}^\alpha+\binom{\alpha}{1}\lambda_{n-1}^{\alpha-1}x+\binom{\alpha}{2}\lambda_{n-1}^{\alpha-2}x^2+\cdots+\binom{\alpha}{\alpha}x^\alpha\right).
\end{eqnarray*}
Clearly, the minimum and maximum powers of $x$ possible in the above product are $0$ and $(n-1)\alpha$, respectively. Therefore, $r-\alpha$ varies from 0 to $(n-1)\alpha$, and any particular value assumed by it in this range has to be the resultant of the powers of $x$ in the factors $(\binom{\alpha}{0}\lambda_j^\alpha+\cdots+\binom{\alpha}{\alpha}x^\alpha)$; $j=1,...,n-1$. As a result, we look for the partitions of $r-\alpha$ using exactly $n-1$ non-negative integers which are less than or equal to $\alpha$, since the power of $x$ varies from 0 to $\alpha$. Moreover, the different orderings of the partition constituents correspond to the exchange of different $\lambda$'s. Since the multidimensional-integral in~(\ref{gnab}) is symmetric under the exchange of eigenvalues, we may focus on a particular ordering and multiply the resultant integral by the suitable combinatorial factor, which for a partition indexed by say $\varphi$, out of $1$ to $L$ in~(\ref{partition}), can be seen to be $(n-1)!/\prod_{k=1}^{l_\varphi}s_{\varphi,k}!$. Furthermore, this factor appears with 
\begin{eqnarray*}
\binom{\alpha}{p_{\varphi,k}}^{s_{\varphi,1}}\binom{\alpha}{p_{\varphi,2}}^{s_{\varphi,2}}\cdots\binom{\alpha}{p_{\varphi,l_\varphi}}^{s_{\varphi,l_\varphi}} x^{s_{\varphi,1}p_{\varphi,1}+s_{\varphi,2}p_{\varphi,2}+\cdots+s_{\varphi,l_\varphi}p_{\varphi,l_\varphi}}\\
~~~\times \lambda_1^{\alpha-\p_{j,1}}\lambda_2^{\alpha-\p_{j,1}}\ldots \lambda_{s_{j,1}}^{\alpha-\p_{j,1}} \lambda_{s_{j,1}+1}^{\alpha-\p_{j,2}}\ldots \lambda_{s_{j,2}}^{\alpha-\p_{j,2}} \ldots \lambda_{n-1}^{\alpha-\p_{l_j}}.
\end{eqnarray*}
The final result~(\ref{coeff}) therefore follows by summing the above over $L$ distinct partitions of $r-\alpha$ and applying the multidimensional-integral appearing in~(\ref{gnab}).

We consider an example to enunciate the above. Suppose $n=5, \alpha=3, \beta=2$, and we are interested in finding the coefficient of $x^r$ with $r=7$. Then we look for the partition of $7-3=4$ and find the unique partitions $\{3,1,0,0\}, \{2,2,0,0\}, \{2,1,1,0\}, \{1,1,1,1\}$ up to ordering. Therefore, we have $L=4$, and the following parameters:
\begin{eqnarray*}
&l_1=3: p_{1,1}=3, s_{1,1}=1; p_{1,2}=1, s_{1,2}=1; p_{1,3}=0, s_{1,3}=2,\\
&l_2=2: p_{2,1}=2, s_{2,1}=2; p_{2,2}=0, s_{2,2}=2,\\
&l_3=3: p_{3,1}=2, s_{3,1}=1; p_{3,2}=1, s_{1,2}=2; p_{3,3}=0, s_{3,3}=1,\\
&l_4=1: p_{4,1}=1, s_{4,1}=4.
\end{eqnarray*} 
Equation~(\ref{coeff}) then tells that the coefficient of $x^7$ would be
\begin{eqnarray*}
\kappa_7=4!\,c_{5,3,2}\Bigg[&&
\frac{\binom{3}{3}^1}{1!}\frac{\binom{3}{1}^1}{1!}\frac{\binom{3}{0}^2}{2!}\expval{\lambda_1^0\lambda_2^2\lambda_3^3\lambda_4^3}_\Lambda
+\frac{\binom{3}{2}^2}{2!}\frac{\binom{3}{0}^2}{2!}\expval{\lambda_1^1\lambda_2^1\lambda_3^3\lambda_4^3}_\Lambda\\
&&+\frac{\binom{3}{2}^1}{1!}\frac{\binom{3}{1}^2}{2!}\frac{\binom{3}{0}^1}{1!}\expval{\lambda_1^1\lambda_2^2\lambda_3^2\lambda_4^3}_\Lambda
+\frac{\binom{3}{1}^4}{4!}\expval{\lambda_1^2\lambda_2^2\lambda_3^2\lambda_4^2}_\Lambda
\Bigg].
\end{eqnarray*} 
We find that
$\expval{\lambda_1^0\lambda_2^2\lambda_3^3\lambda_4^3}_\Lambda=3175200$,
$\expval{\lambda_1^1\lambda_2^1\lambda_3^3\lambda_4^3}_\Lambda=1360800$,
$\expval{\lambda_1^1\lambda_2^2\lambda_3^2\lambda_4^3}_\Lambda=680400$,
$\expval{\lambda_1^2\lambda_2^2\lambda_3^2\lambda_4^2}_\Lambda=302400,$
which gives $\kappa_7=159/16$. This agrees with the coefficient of $x^7$ extracted after applying the recursion, as it should.

\section{Mathematica Codes}
\label{AppMtmk}
The following code can be implemented in Mathematica~\cite{Mathematica} to obtain exact expressions for the smallest eigenvalue density for the unrestricted $\beta$-Wishart-Laguerre ensemble:
 \begin{figure*}[ht!]
\includegraphics[width=1.1\textwidth]{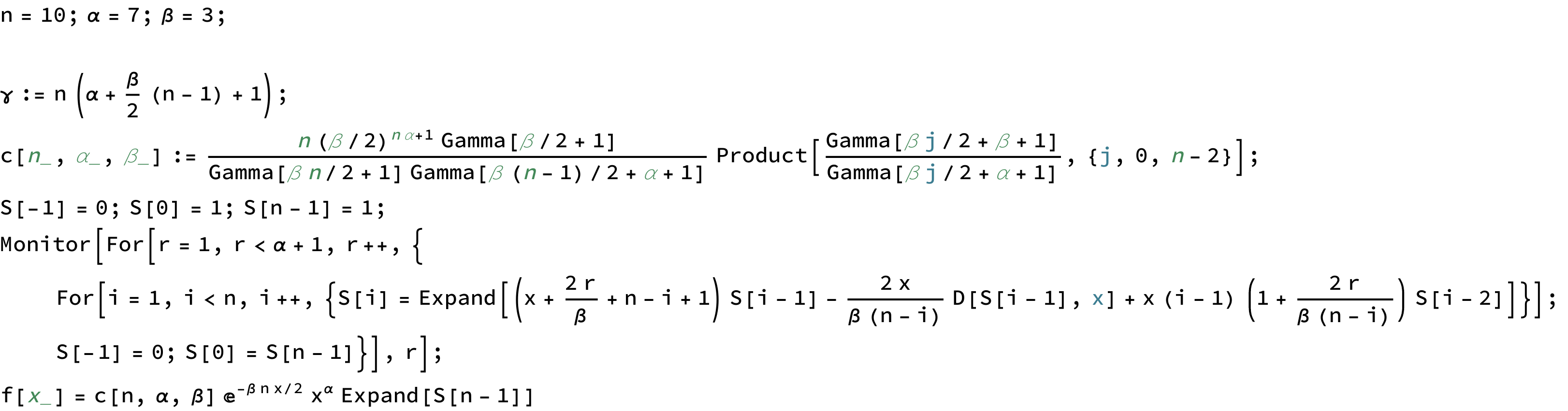}
 \label{code1}
\end{figure*}\\
For generating the smallest eigenvalue density for unit-trace $\beta$-Wishart-Laguerre ensemble, the following code can be used along with the above.
 \begin{figure*}[ht!]
\includegraphics[width=0.8\textwidth]{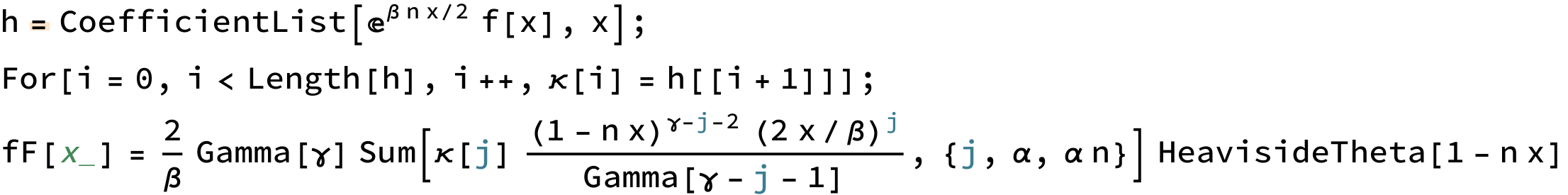}
 \label{code2}
\end{figure*}\\

\noindent
Subsequently, the following codes can be used to obtain the moments:
 \begin{figure*}[ht!]
\includegraphics[width=0.55\textwidth]{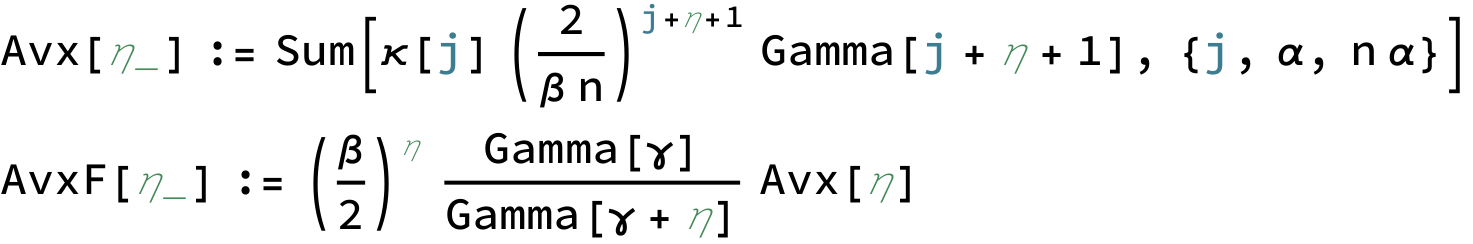}
 \label{code3}
\end{figure*}

\newpage


\end{document}